# Classification of seizure and seizure-free EEG signals based on empirical wavelet transform and phase space reconstruction

Hesam Akbari
Department of Biomedical Engineering, South Tehran Branch, Islamic Azad University, Tehran, Iran
st_h.akbari@azad.ac.ir

Somayeh Saraf Esmaili
Department of Biomedical Engineering, Garmsar Branch, Islamic Azad University, Garmsar, Iran
s.sesmaeily@iau-garmsar.ac.ir

Sima Farzollah Zadeh
Department of Biomedical Engineering, Science and Research Branch, Islamic Azad University, Tehran, Iran
s.farzollahzadeh@gmail.com

*Abstract*— Epilepsy is a brain disorder due to abnormal activity of neurons and recording of seizures is of primary interest in the evaluation of epileptic patients. A seizure is the phenomenon of rhythmicity discharge from either a local area or the whole brain and the individual behavior usually lasts from seconds to minutes. In this work, empirical wavelet transform (EWT) is applied to decompose signals into Electroencephalography (EEG) rhythms. EEG signals are separated to delta, theta, alpha, beta and gamma rhythms using EWT. The proposed method has been evaluated by benchmark dataset which is freely downloadable from Bonn University website. 95% confident ellipse area is computed from 2D projection of reconstructed phase space (RPS) of rhythms as features and fed to K-nearest neighbor classifier for detection of seizure (S) and seizure free (SF) EEG signals. Our proposed method archived 98% accuracy in classification of S and SF EEG signals with a tenfold cross-validation strategy that is higher than previous techniques.

## I. INTRODUCTION

Epilepsy is a brain disorder due to abnormal activity of neurons. About 50 million people suffering from epilepsy who most of them are living in developing countries [1].One of the commonly used electrophysiological monitoring methods to detect of epilepsy is Electroencephalography (*EEG*). Epileptic seizures in human brain frequently manifest spikes in EEG signals which can be analyzed visually by the experts [2]. Visual inspection of long EEG records to detect presence of epileptic seizures can be cumbersome and time consuming activity. Therefore, an automatic method to detection and classification of normal and epilepsy seizure is desirable. Recently, many methods have been developed for this proposes. Classification of seizure (S) and seizure free (SF) epilepsy EEG signals using permutation entropy have been reported in [3]. In [4], Mean degree and the mean strength of horizontal visibility graph (HVG) have been used in k-nearest neighbor classifier for detecting S EEG signals. Classification of S and SF EEG signals using clustering technique and support vector machine (SVM) have been reported in [5]. In [6], linear prediction error energy feature has been used for classification of S and SF EEG signals. In [7], fractional linear prediction error (FLP) and signal energy are used as features for classification S and SF EEG signals. Empirical mode decomposition (EMD) has been proposed to decompose an input signal into intrinsic mode functions (IMFs) [8]. In [9], 95% confident area measure of second-order difference plot (SODP) from IMFs is extracted and artificial neural network (ANN) classifier is applied to classify EEG signals in S and SF groups. In [10], dual tree complex wavelet transform (DTCWT) decomposed EEG signals and various entropies and statistically based features computed as input to general regression neural network (GRNN) classifier for discriminate S and SF EEG signals. Also, Classification of S and SF EEG signals using tunable-Q wavelet transform (TQWT) and Kraskov entropy has been reported in [11]. Recently, empirical wavelet transform (EWT) has been proposed to analyze non-stationary signals [12]. It decomposes input signal based on adaptive filter bank. Band pass of EWT filters is determined using significant segmentation of spectrum of the input signal. In this work, EWT is applied to decompose signals into EEG rhythms. Traditional wavelet transform and EMD separates these rhythms by multiple number decompositions, while EWT can extract these rhythms by one-step proses. The reconstructed phase space (RPS) can manifest the dynamic of chaotic. Due to the chaotic nature of the EEG signals, RPS is applied to exhibit of EEG rhythms in 2D projection. In this work, 95% confident ellipse area is computed from 2D projection of RPS of rhythms as features and fed to K-nearest neighbor classifier by tenfold cross validation strategy for detection of S and SF EEG signals.

## II. PROPOSED METHOD

In this paper, EEG signals are separated to delta ($\delta$), theta ($\theta$), alpha ($\alpha$), beta ($\beta$) and gamma ($\gamma$) rhythms using EWT. Then 2D projections of rhythms are plotted by RPS. Finally, 95% confident ellipse area of rhythms computed as input features to KNN classifier. Fig.1 shows the block diagram of the proposed method.

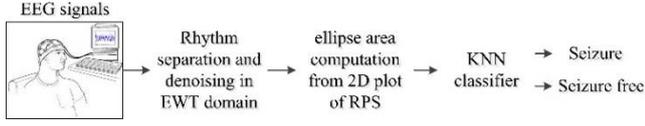

Figure.1 Proposed method.

A. *Database used*

The Proposed method has been evaluated by benchmark dataset which is freely downloadable from Bonn University website [13]. This database consists of 5 subset called A, B, C, D and E, that each subset has 100 EEG signal were sampled at a rate of 173.61 Hz. The duration of each EEG signal is 23.6 second, so has 4096 samples. The subset of A and B were recorded from five healthy subjects in eyes opened and closed condition, respectively. The subset of C and D were recorded from five patients who had completely recovered from seizure control after surgery of epileptic locations. The subset E is composed of EEG signals with epileptic seizure activities that are observed in epileptogenic zone. In this work, signals in C and D subsets are considered as SF EEG signals, and signals in E subset are considered as S EEG signals. Fig. 2 shows an S and SF signal.

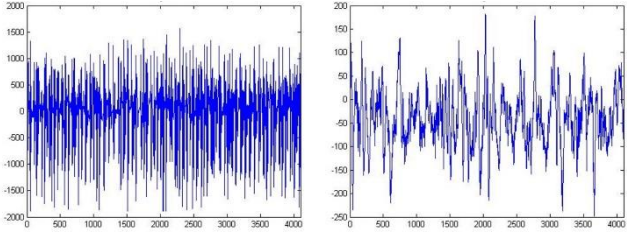

Figure.2 Left and right shows S and SF signal.

B. *Empirical wavelet transform (EWT)*

EWT decompose signal by generating adaptive filter bank corresponding to the input signal spectrum. Band pass of adaptive filter bank is determinate using proper segmentation of the spectrum. In EWT toolbox, many methods are proposed to proper segmentation of the spectrum [14]. Segmentation of spectrum to [0-4 Hz], [4-8 Hz], [8-16 Hz], [16-30] and [30-60] bands will be resulted $\delta$, $\theta$, $\alpha$, $\beta$ and $\gamma$ rhythms, respectively. For this propose, we set the cut-off frequencies as $f_{cut} = \{4, 8, 16, 30, 60\}$ and use them for construct scaling function and wavelet functions. Filters of scaling function $\varphi(\omega_f)$ and wavelet functions $\psi(\omega_f)$ are constructed in Fourier domain based on little wood-Paley and Meyer wavelets as follows [13]:

$$\varphi(\omega_f) = \begin{cases} 1 & \text{if } |\omega_f| \leq (1-\lambda)f_1 \\ \cos(\frac{\pi\beta(\lambda, f_1)}{2}) & \text{if } (1-\lambda)f_1 \leq |\omega_f| \leq (1+\lambda)f_1 \\ 0 & \text{otherwise} \end{cases} \quad (1)$$

$$\psi_{i=1,2,\ldots 5}(\omega_f) = \begin{cases} 1 & \text{if } (1+\lambda)f_i \leq |\omega_f| \leq (1-\lambda)f_{i+1} \\ \cos(\frac{\pi\beta(\lambda, f_{i+1})}{2}) & \text{if } (1-\lambda)f_{i+1} \leq |\omega_f| \leq (1+\lambda)f_{i+1} \\ \sin(\frac{\pi\beta(\lambda, f_i)}{2}) & \text{if } (1+\lambda)f_i \leq |\omega_f| \leq (1+\lambda)f_i \\ 0 & \text{otherwise} \end{cases} \quad (2)$$

Where $\beta(\lambda, \omega_i) = \beta(\frac{|\omega_f| - (1-\lambda)}{2\lambda\omega_f})$ and $\lambda$ Parameter defined as $\lambda < \min(\frac{\omega_{i+1} - \omega_i}{\omega_{i-1} + \omega_i})$, make sure that the EWT coefficients are in $L^2(\Re)$ space.

In this work, $\lambda$ parameter is computed to 0.1825. Also, $\beta(y)$ is arbitrary function defined as:

$$\beta(y) = \begin{cases} 0 & \text{if } y \leq 0 \\ \beta(y) + \beta(1-y) = 1 & \forall y \in [0,1] \\ 1 & \text{if } y \geq 1 \end{cases} \quad (3)$$

Finally, each rhythm can be found by the inner product of EEG signals with corresponding filters.

The EWT filter bank generates tight frame, transition band of the filters is very small, and pass-band and stop-band ripples are negligible in the band-pass filters [13]. As a result, separated rhythms will have very less aliasing, which leads to more precise representation.

Fig.3 shows the EWT filter bank for EEG rhythms separation and Fig.4 shows the separated rhythms from S and SF signals.

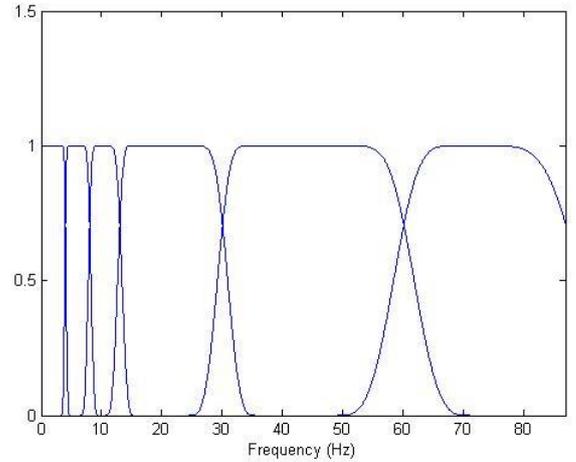

Figure. 3 EWT filter bank.

It should be noted that the any frequency content greater than the highest frequency range of gamma is discarded as noise signal. EWT has been used previously for detection of focal EEG signals [15].

C. *The reconstructed phase space and feature extraction*

The reconstructed phase space (RPS) uses to show the nonlinear nature of the stabilogram signal [16]. In this paper, RPS of rhythms is used as a visual image for evaluation of the dynamical behavior of S and SF EEG signals. RPS has been used previously for classification of patient independent heartbeat [17]. The RPS generation requires determination of delay time $\tau$ and embedding dimension $d$ which can be obtained by mutual information (MI) [18, 19] and the nearest neighbor (NN) method [18], respectively.

For the signal $V = \{v_1, v_2, v_3, ... v_k\}$, where K is the total number of data point, the RPS defined as:

$$Y_k = (V_k, V_{k+\tau}, ..., V_{k+(d-1)\tau}) \quad (3)$$

Where,

$$k = 1, 2, ..., K - (d-1)\tau \quad (4)$$

In this work, $\tau$ and $d$ values are chosen empirically to 1 and 2, respectively. The 2D projection of RPS of a signal obtained by plotting $V_k$ against $V_{k+1}$. 2D RPS of rhythms of S and SF signals are show in figure 5.

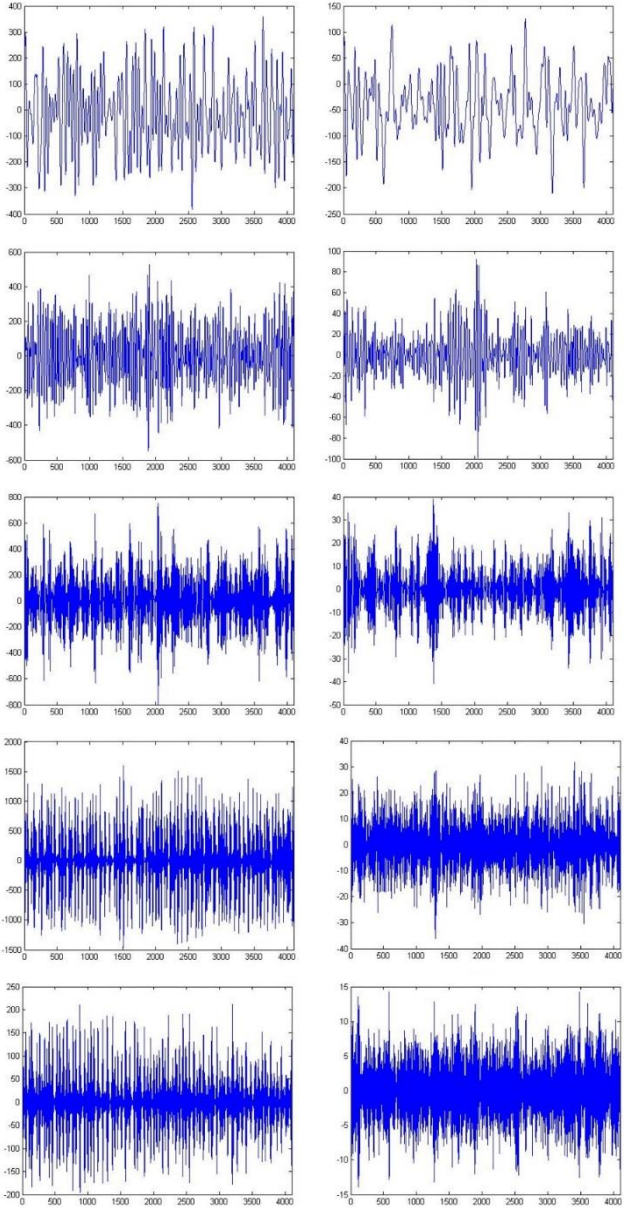

Figure. 4 From up to down of left and right columns are separated $\delta$, $\theta$, $\alpha$, $\beta$ and $\gamma$ rhythms for S and SF signal.

It is clear form figure 5 that the 2D RPS of rhythms have elliptical patterns. Besides, S signals occupy more are in 2D RPS. It motivates us to compute the ellipse area of 2D RPS of rhythms for classification of S and SF EEG signals.

The procedure to calculate the 95% confidence ellipse area from the RPS can be given as [20, 21]:

Compute the mean values of $V_k$ and $V_{k+1}$ as:

$$S_X = \sqrt{\frac{1}{K-1}\sum_{k=1}^{K-1} V_k^2} \quad (5)$$

$$S_Y = \sqrt{\frac{1}{K-1}\sum_{k=1}^{K-1} V_k^2} \quad (6)$$

$$S_{XY} = \frac{1}{K-1}\sum_{k=1}^{K-1} V_k V_{k+1} \quad (7)$$

Compute C parameter as:

$$C = \sqrt{(S_X^2 + S_Y^2) - 4(S_X^2 S_Y^2 - S_{XY}^2)} \quad (8)$$

$$a = 1.7321\sqrt{(S_X^2 + S_Y^2 + C)} \quad (9)$$

$$b = 1.7321\sqrt{(S_X^2 + S_Y^2 - C)} \quad (10)$$

From the parameters 'a' and 'b', the ellipse area is computed as equation (11):

$$A_{ellipse} = \pi a b \quad (11)$$

*D. K nearest neighbor (KNN) classifier*

KNN is supervised classifier with very easy theory and implementation. The KNN classifies any sample of test data considering their K closed neighbor samples in the train data. Test samples belong to the group which has more members among K closed neighbor. Distance computation method and number of K are two parameters of KNN classifier. In this work, Euclidean and city block distances are used by varied number of k from 1 to 10 to classify S and SF EEG signals. The sensitivity (SEN), specificity (SPE), accuracy (ACC), positive predictive value (PPV) and negative predictive value (NPV) parameters [22] are computed to evaluate the classifier performance. KNN classifier is used previously for classification of focal and non-focal EEG signals [23] [24] and classification of depression patients and normal subjects EEG signals [25].

III. RESULTS AND DISCUSSION

In this paper, we propose a method based on extracted rhythms in EWT domain and RPS to classification of S and SF EEG signals. EEG signals are decomposed to EEG rhythms using EWT. Then, 2D RPS of rhythms is plotted and 95% of elliptical pattern computed as feature. The Kruskal–Wallis statistical test evaluated the features corresponding to their p-values. The features with less than 0.05 p-value is selected as significant features [26]. P-values of extracted features corresponding to each rhythms are written in Table 1. It is evidence that all rhythms show good discrimination between S and SF EEG signals (p<0.05). Although, we could use from all of the computed features, but in order to reduce the complexity of the classifier, the size of the features set are selected to 2. In other words, extracted features from in any time two rhythms are used in the classifier. extracted features from both $\delta$-$\theta$, $\delta$-$\alpha$, $\delta$-$\beta$, $\delta$-$\gamma$, $\theta$-$\alpha$, $\theta$-$\beta$, $\theta$-$\gamma$, $\alpha$-$\beta$, $\alpha$-$\gamma$ and $\beta$-$\gamma$ rhythms are fed to a KNN classifier with the Euclidean and city-block distance in tenfold cross-validation strategy. The Cross-validation technique is applied to ensuring the reliable classification performance [27].

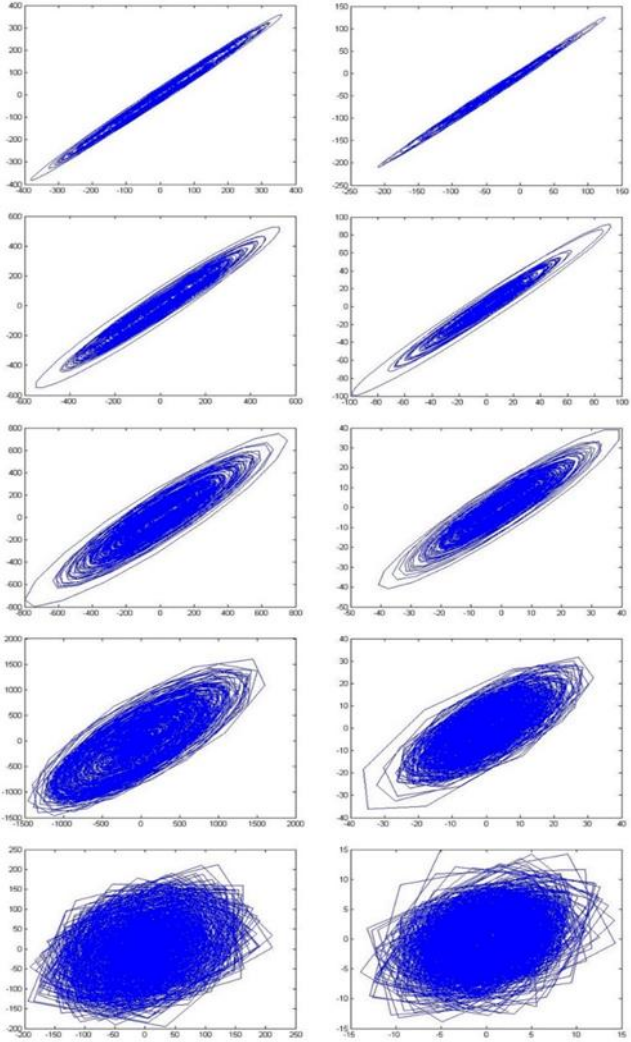

Figure. 5 From up to down shows the 2D RPS of $\delta$, $\theta$, $\alpha$, $\beta$ and $\gamma$ rhythms for S (left) and SF (right) signal.

It is clear from these two tables that, $\alpha$-$\beta$ and $\alpha$-$\gamma$ features set using KNN classifier with Euclidean distance and $\alpha$-$\beta$ features set using KNN classifier with city-block distance showed a good ACC of 98%. We have compared our proposed method with existing methods studies on the same database in Table 4. It is clearly evident that the proposed method archived highest classification accuracy. In [11], researchers have used EMD and SODP to detecting S and SF EEG signals. Although they reported the best accuracy of 97.75%, which are very close to the highest accuracy (98%) in our method, they are used EMD which suffers from the mode-mixing problem and noise sensitive. In even that, our proposed method not noise sensitive.

TABLE I. Computed p-values for features.

| Rhythms | delta | theta | alpha | beta | gamma |
|---|---|---|---|---|---|
| p-value | $4.83 \times 10^{-25}$ | $3.31 \times 10^{-43}$ | $5.64 \times 10^{-44}$ | $2.88 \times 10^{-44}$ | $1.11 \times 10^{-40}$ |

TABLE II. Performance of KNN classifier with cityblock distance.

| Feature set | ACC (%) | SEN (%) | SPE (%) | PPV (%) | NPV (%) | Number of K |
|---|---|---|---|---|---|---|
| $\delta$-$\theta$ | 96 | 95 | 96.50 | 93.84 | 97.58 | 2 |
| $\delta$-$\alpha$ | 95 | 93 | 96 | 92.27 | 96.49 | 2 |
| $\delta$-$\beta$ | 92 | 84 | 96 | 91.82 | 92.77 | 6 |
| $\delta$-$\gamma$ | 91.66 | 84 | 95.50 | 91.38 | 92.44 | 2 |
| $\theta$-$\alpha$ | 97.66 | 97 | 98 | 96 | 98 | 5 |
| $\theta$-$\beta$ | 97.33 | 96 | 98 | 96.69 | 98.11 | 7 |
| $\theta$-$\gamma$ | 95.66 | 91 | 98 | 95.95 | 95.75 | 8 |
| $\alpha$-$\beta$ | 98 | 98 | 98 | 96.18 | 99 | 6 |
| $\alpha$-$\gamma$ | 98 | 98 | 98 | 96.36 | 99.04 | 8 |
| $\beta$-$\gamma$ | 97 | 95 | 98 | 96.36 | 97.51 | 3 |

Table III. Performance of KNN classifier with Euclidean distance.

| Feature set | ACC (%) | SEN (%) | SPE (%) | PPV (%) | NPV (%) | Number of k |
|---|---|---|---|---|---|---|
| $\delta$-$\theta$ | 97 | 96 | 97.5 | 95.27 | 98.04 | 4 |
| $\delta$-$\alpha$ | 94.33 | 90 | 96.50 | 93 | 95 | 2 |
| $\delta$-$\beta$ | 91.66 | 82 | 96.50 | 92.07 | 91.67 | 3 |
| $\delta$-$\gamma$ | 91 | 80 | 96.50 | 93.34 | 90.79 | 6 |
| $\theta$-$\alpha$ | 97.66 | 97 | 98 | 96.27 | 98 | 6 |
| $\theta$-$\beta$ | 97.33 | 96 | 98 | 96.18 | 98.04 | 9 |
| $\theta$-$\gamma$ | 95.66 | 92 | 97.50 | 94.95 | 96.21 | 4 |
| $\alpha$-$\beta$ | 98 | 98 | 98 | 96.27 | 99.02 | 3 |
| $\alpha$-$\gamma$ | 97.66 | 97 | 98 | 96.27 | 98.54 | 9 |
| $\beta$-$\gamma$ | 97 | 96 | 97.50 | 95 | 98.21 | 7 |

## IV. CONCLUSION

Epileptic seizures in human brain frequently manifest spikes in EEG signals [2] which can be analyzed visually by the experts. Visual inspection of long EEG records to detect presence of epileptic seizures can be cumbersome and time consuming activity. This paper proposed a method based on rhythm separation using EWT for classification the S and SF EEG signals. 95% of the elliptical pattern of 2D projection of RPS is used on rhythms of EEG. Statistically significant features (p-values< 0.05) are chosen from Kruskal–Wallis statistical testing. Proposed method archived 98% classification accuracy to the detecting of S EEG signals using KNN classifier with city-block and Euclidean distances. The proposed method archived highest classification accuracy compared with previous techniques. Our proposed method measured the 2D area of rhythms in as a feature. It does not compute the degree of variability or complexity of rhythms, which can be a good parameter to the classification of EEG signals in two groups' of S and SF. Recently, variation mode decomposition (VMD) has been propose instead of EWT method [28]. In future, the performance of the proposed feature with VMD will be evaluated.

TABLE IV. Comparison of proposed method with the exiting work.

| Reference | Method | Set | ACC (%) |
|---|---|---|---|
| [3] (2012) | Permutation entropy and SVM classifier | N and S<br>F and S | 88.83<br>83.13 |
| [4] (2014) | Degree and strength of HVG and KNN classifier | N and S<br>F and S | 98<br>93 |
| [5] (2011) | Clustering and SVM classifier | N and S<br>F and S | 97.69<br>93.91 |
| [6] (2010) | Linear prediction error energy | N, F and S | 94 |
| [7] (2014) | FLP error energy and signal energy and SVM classifier | N, F and S | 95.33 |
| [9] (2014) | 95% Confidence area measure of SODP of IMFs and ANN classifier | N, F and S | 97.75 |
| [10] (2016) | DTCWT and GRNN | N, F and S | 95.15 |
| [11] (2017) | TQWT and Korsakov entropy and Least square SVM | N, F and S | 97.5 |
| Proposed method | 95% Confidence area measure of RPS of rhythms in EWT domain and KNN classifier | N, F and S | 98 |